# Benefits of current percolation in superconducting coated conductors

**N.A. Rutter, J.H. Durrell, M.G. Blamire and J.L. MacManus-Driscoll**

*Department of Materials Science & Metallurgy, Pembroke St., Cambridge, CB2 3QZ, UK*

**H. Wang and S.R. Foltyn**

*Superconductivity Technology Center, Los Alamos National Laboratory, Los Alamos, NM 87545, USA*

The critical currents of MOD/RABiTS and PLD/IBAD coated conductors have been measured as a function of magnetic field orientation and compared to films grown on single crystal substrates. By varying the orientation of magnetic field applied in the plane of the film, we are able to determine the extent to which current flow in each type of conductor is percolative. Standard MOD/RABiTS conductors have also been compared to samples whose grain boundaries have been doped by diffusing Ca from an overlayer. We find that undoped MOD/RABiTS tapes have a less anisotropic in-plane field dependence than PLD/IBAD tapes and that the uniformity of critical current as a function of in-plane field angle is greater for MOD/RABiTS samples doped with Ca.

Coated conductor YBa$_2$Cu$_3$O$_{7-\delta}$ (YBCO) tapes fabricated by both the Ion Beam Assisted Deposition (IBAD) and Rolling Assisted Biaxially Textured Substrates (RABiTS) techniques are now capable of carrying current densities ($J$) [1] in excess of 3 MAcm$^{-2}$ at 77 K in self field. This approaches the value measured in films grown on single crystal substrates. For IBAD tapes, this can be explained by the fact that the texture achieved in these samples can be described as "single-crystal-like", the FWHM of x-ray rocking curves and Φ scans being around 2-3° [2]. RABiTS conductors however usually have less sharp textures [3] and yet still have high values of J$_c$, higher than would be expected based upon models of the grain boundary networks [4]. Recent work [5] has indicated that this may be due to the fact that the grain boundaries in thick films grown using techniques in which a transient liquid phase may be present can be highly overgrown and meandering, meaning that their effective cross-sectional area will be much greater than an equivalent planar boundary.

Depending upon the conditions of temperature and applied magnetic field, the critical current of a coated conductor may either be limited by the in-grain properties, the grain-boundary $j_c$ or some combination of the two. Measurements made with the magnetic field applied parallel to the c-axis of the superconductor [6] have shown that a crossover between the two regimes can be identified such that at low magnetic fields, $J_c$ is limited by the grain boundaries, but at high fields the properties of the grains determine the critical current [7].

This type of result is expected based on measurements made on films grown on single crystalline and bi-crystalline substrates. In self field, $j_c$ for a single [001]-tilt boundary with a misorientation angle greater than around 2° is lower than that of the neighbouring grains [8]. However, application of a magnetic field suppresses the intra-granular critical current more than it reduces the grain boundary critical current such that there is a crossover field above which the grains limit the critical current. The magnitude of this crossover field depends on the boundary misorientation angle and is highest when the grains are highly misoriented.

In bi-crystal boundaries a crossover between boundary and grain dominated behaviour has also been identified for fields oriented within the plane of the superconducting film. Importantly, for a given field magnitude and temperature, a crossover can be identified as a function of the in-plane orientation such that the boundary becomes the limiting factor when the field is aligned within some angle of it [9].

In addition to giving information regarding the separate effects of grains and boundaries, the effect on $J_c$ of an in-plane field is also of practical importance for coated conductors. One application of such tapes is in winding magnets, in which case the field will primarily be oriented in the plane and perpendicular to the current-flow direction.

It has been shown that the uniform addition of calcium to YBCO can lead to an increase in $j_c$ at low temperatures for low-angle grain boundaries [10]. The high quality coated conductor tapes which are currently being produced have sharp x-ray rocking curves and the majority of grains are misoriented from their neighbours by just a few degrees. The disadvantage of uniformly doping YBCO with Ca is that it tends to reduce the critical temperature ($T_c$) of the grains, so it would be ideal to dope the grain boundaries preferentially. One way in which this may be achieved is by exploiting the fact that the diffusion rate along a grain boundary in a material is significantly greater than that within a grain. Such a technique has been studied by Berenov et al. [11] with the calcium source being a $Ca(NO_3)_2$ solution. SIMS studies showed that diffusion of Ca in c-axis films is as much as 2 orders of magnitude faster along the grain boundary.

The RABiTS coated conductor tapes used in this study were fabricated by American Superconductor Corporation's MOD process with an architecture of YBCO/$CeO_2$/YSZ/$Y_2O_3$/Ni-W and YBCO thickness around 0.8 μm. The IBAD tape was fabricated at Los Alamos National Laboratory with the YBCO layer being grown by PLD on an IBAD MgO layer and $CeO_2$ cap and was also around 0.8 μm thick. Samples 13 mm in length and 5 mm in width were cut such that the long dimension, along which the current flows, was parallel to the rolling direction of the RABiTS tape. The single crystalline film was grown on a $SrTiO_3$ substrate by PLD and was around 1 μm thick. For the samples which were Ca doped, a $CaZrO_3$ film, approximately 100 nm thick, was deposited directly on top of the YBCO layer by pulsed laser deposition from a $CaZrO_3$ target in 0.2 mbar $O_2$. $CaZrO_3$ was chosen as the calcium source as it is a stable phase which is not sensitive to the presence of $H_2O$ or $CO_2$, unlike many other Ca compounds. In addition, Zr does not substitute into the YBCO lattice. Subsequently, the samples were annealed at 500°C in flowing oxygen (99.999%) at 1 atm for 30 minutes in order to diffuse Ca along the grain boundaries.

X-ray rocking curves were measured using a vertical diffractometer in order to quantify the out-of plane alignment of the coated conductor samples. Pole figures and phi scans were also carried out using a 4-circle goniometer to confirm the in-plane crystallographic texture. $T_c$ measurements were performed using a dip probe and liquid helium. Using a 4-terminal transport measurement, the resistance can be measured from 300 K down to 4.2 K and the superconducting transition temperature and transition width determined. In order to facilitate measurement of the coated conductor tapes, 125 μm wide current tracks were patterned using photolithography and chemical etching with 2.5% phosphoric acid. Critical current measurements were made as a function of temperature and the magnitude and angle of aligned magnetic field using a two-axis goniometer mounted in an Oxford Instruments 8 T magnet [12]. Values of the critical current were extracted from the I-V characteristics using a voltage criterion of 1 μV.

Figure 1a) shows X-Ray rocking curves (ω-scans) of the (005) peaks for the films. The film grown on a single crystal has a very sharp out-of plane texture, the PLD-IBAD tape is only slightly more misaligned whilst the YBCO grown by MOD on a RABiTS tape shows a significantly broader rocking curve. The FWHM values are noted in table I. Figure 1b) shows the (115) Φ-scans and the FWHM values are also included in table I. The degree of in-plane alignment of the various samples shows a similar trend to the out-of-plane texture. The texture in the YBCO layer is not affected by deposition of $CaZrO_3$ and subsequent annealing.

The critical temperature of the undoped MOD/RABiTS samples is very high, around 94 K. This is only reduced by around 1-2 K by the Ca-doping procedure. $T_c$ for the PLD/IBAD tapes is 92 K and that of the film grown on single crystal 90 K. Table I shows the actual critical current densities measured in the samples at 80 K and 1 T.

Figure 2 shows the dependence of $J_c$ on the in-plane orientation of magnetic field for a single crystalline film, PLD/IBAD tape, undoped MOD/RABiTS tape and Ca-doped MOD/RABiTS tape. The different types of sample show very different behaviour as the field is rotated in the plane of the tape between the force-free (FF) orientation (ϕ=0°) parallel to the current flow, and the transverse direction (ϕ=90°).

Consider first the single-crystalline film. The shape of the ϕ-scan is due to the variation in Lorentz force. When the current direction is aligned with the magnetic field (ϕ=0°), the force acting on the flux lines is less than for field applied at ϕ=90°. For the coated conductors, we first note that the crossover between grain and boundary limited behaviour that is observed in bi-crystal samples [8] was not seen in the coated conductor tapes. We attribute this to the fact that in the coated conductors the field is never aligned along all the relevant grain boundary segments. In addition it is evident that the tape $J_c$ shows significantly less dependence on in-plane field angle than the single crystalline film. At 1 T and 80 K, the critical current when the field is aligned transverse to the current direction is less than half that in the FF geometry. We will use the symbol $\zeta$ to represent the ratio $J_c(\phi=90°)/J_c(\phi=0°)$. This value can be seen from fig. 2 to be around 0.59 for the PLD/IBAD tape, 0.8 for the undoped MOD/RABiTS sample and 0.85 for the doped tape. We have measured $\phi$-scans at a range of fields in order to see the variation of $\zeta$ as shown in fig 3.

The very different levels of in-plane anisotropy of the various types of sample can be explained by considering how the current flows at the microscopic level. We assume that in the single crystalline film, all the current flows uniformly parallel to the macroscopic transport direction. Hence the field aligns fully parallel with the current when $\phi=0°$ and fully perpendicular to the current when $\phi=90°$, leading to a large force free effect. We know however that in coated conductor tapes the grain boundaries cause percolation at a microscopic level such that the current has to cross perpendicular to grain boundaries. Hence the field is never completely parallel to, nor completely transverse to all the current. Because the grain boundaries in MOD/RABiTS have higher misorientation angles than in PLD/IBAD, we would expect more percolation and hence a greater flattening of the $\phi$-scan, as is observed.

The Ca-doped tapes produce even flatter curves suggesting that the doping process produces more percolation. This is contrary to what might be expected as the Ca is supposed to improve the properties of the grain boundaries, such that they have less effect. It is likely that overlayer doping leads to a non-uniform doping profile through the sample thickness [13], causing a greater non-uniformity of current flow than in an undoped MOD/RABiTS tape.

In summary, when a magnetic field is rotated within the plane of the film, the critical current of coated conductors shows less variation with angle than that of single crystalline films. Further, MOD/RABiTS tapes are more isotropic than PLD/IBAD conductors in this regard. This is likely to be due to the fact that the crystallographic texture in these multigranular tapes produces non-uniform current flow (percolation). Doping an MOD/RABiTS tape with Ca by diffusing from an overlayer flattens the $\phi$-scans further, without changing the crystallographic texture. This is probably due to the non-uniform nature of the Ca doping. Non-uniformity of the current flow at a microscopic level in coated conductors is a beneficial effect as it makes their performance less dependent of the orientation of the magnetic field. Hence the findings of this work are important for applications of coated conductors in magnet windings.

The authors acknowledge the Engineering and Physical Science Research Council who provided funding for this research. We would also like to thank American Superconductor Corporation for providing coated conductor samples and Dr. M.W. Rupich for helpful discussion and suggestions.

# References

1. The uppercase symbol J is used to represent a macroscopic current density whilst the lowercase j represents the local microscopic current density within a grain or crossing a grain boundary.

2. S.R. Foltyn, P.N. Arendt, Q.X. Jia, H. Wang H, J.L. MacManus-Driscoll, S. Kreiskott, R.F. DePaula, L. Stan, J.R. Groves and P.C. Dowden, Appl. Phys. Lett. **82,** 25 (2003).

3. A. Goyal, M. Paranthaman and U. Schoop, MRS Bulletin **29,** 8 (2004).

4. N.A. Rutter and A. Goyal, in High Temperature Superconductivity I : Materials (Springer, Berlin) p. 377-400 (2003).

5. D.M. Feldmann, N.A. Nelson, S.I. Kim, D.C. Larbalestier, T.G. Holesinger, C. Cantoni, and D. T. Verebelyi. J. Mater. Res. Accepted for publication.

6. L. Fernandez, B. Holzapfel, F. Schindler, B. de Boer, A. Attenberger, J. Hanisch, and L. Schultz, Phys. Rev. B, **67**, 052503, (2003).

7. S. I. Kim, D. M. Feldmann, D. T. Verebelyi, C. Thieme, X. Li, A. A. Polyanskii, and D. C. Larbalestier, Phys. Rev. B, **71**, 104501 (2005).

8. D.T. Verebelyi, D.K. Christen, R. Feenstra, C. Cantoni, A. Goyal, D.F. Lee, M. Paranthaman, P.N. Arendt, R.F. DePaula, J.R. Groves, and C. Prouteau, Appl. Phys. Lett., **76** 13 (2000).

9. J.H. Durrell, M.J. Hogg, F. Kahlmann, Z.H. Barber, M.G. Blamire, and J.E. Evetts, "Critical Current of YBa$_2$Cu$_3$O$_{7-\delta}$ Low Angle Grain Boundaries," Phys. Rev. Lett., **90**, 247006 (2003).

10. G.A. Daniels, A. Gurevich, and D. Larbalestier, Appl. Phys. Lett., **77**, 20, (2000).

11. A. Berenov, S.R. Foltyn, C.W. Schneider, P.A. Warburton, and J.L. MacManus-Driscoll, Solid State Ionics, **164** (2003).

12. R. Hertzog and J.E. Evetts, Rev. Sci. Instrum. **65**, 3574, (1994).

13. N.A. Rutter, J.H. Durrell, S.H. Mennema, M.G. Blamire and J.L. MacManus-Driscoll, IEEE Trans. Appl. Supercond., Accepted for publication

**Figure Captions**

Figure 1 a) X-Ray (005) rocking curves and b) (115) Φ-scans showing the differences in crystallographic texture of YBCO films grown on different substrate types

Figure 2. Variation of $J_c$ with in-plane field angle ($\phi$) for T=80 K and $\mu_0H$=1 T. Data is normalised with respect to field oriented in the macroscopic transport direction ($\phi$=0°)

Figure 3. Variation of $\zeta$ as a function of magnitude of applied magnetic field at T=80 K

**Tables**

|  | $\delta\omega(°)$ | $\delta\Phi(°)$ | $J_c$ ($\phi=0°$)/$10^9$ Am$^{-2}$ | $J_c$ ($\phi=90°$)/$10^9$ Am$^{-2}$ |
|---|---|---|---|---|
| Single crystal | 0.5 | 2.4 | 3.0 | 1.3 |
| IBAD | 1.0 | 3.2 | 5.0 | 3.8 |
| Undoped RABiTS | 6.0 | 6.0 | 6.6 | 5.3 |
| Doped RABiTS | 6.0 | 6.0 | 7.0 | 6.0 |

Table I. Values of X-Ray FWHM values and $J_c$ measured for different conductors at 1 T and 80 K.

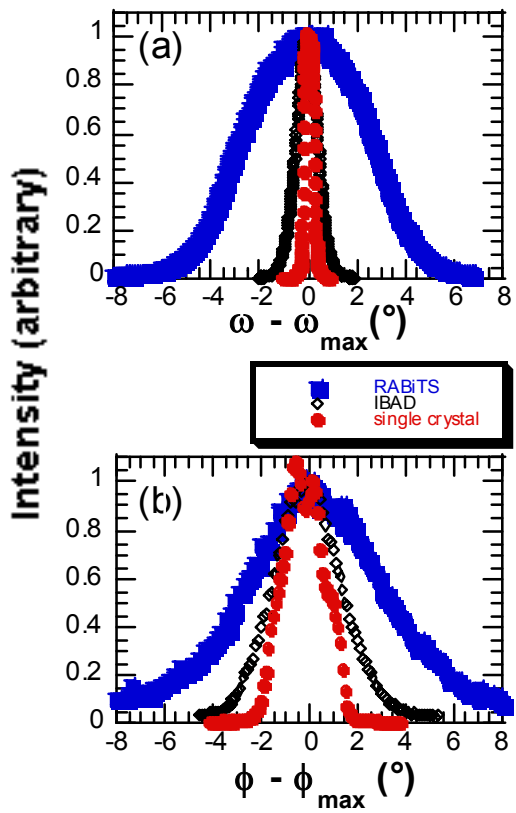

Figure 1

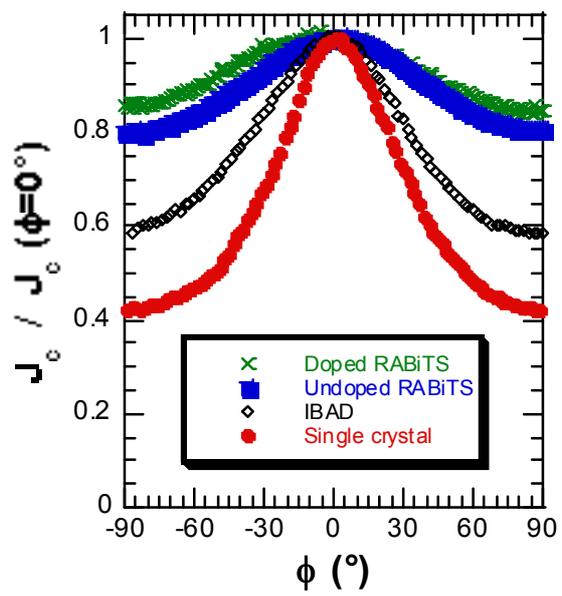

Figure 2

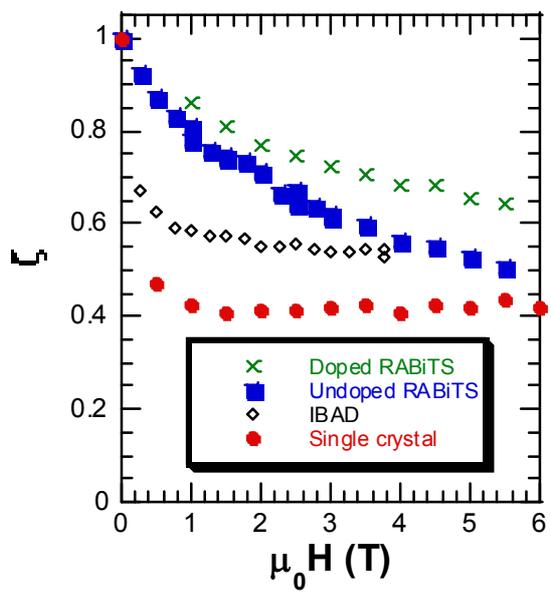

Figure 3